\documentclass[conference]{IEEEtran}
\IEEEoverridecommandlockouts
\usepackage{cite}
\usepackage{amsmath,amssymb,amsfonts}
\usepackage{algorithmic}
\usepackage{graphicx}
\usepackage{textcomp}
\usepackage{xcolor}
\usepackage{subfig}
\usepackage{float}

\def\BibTeX{{\rm B\kern-.05em{\sc i\kern-.025em b}\kern-.08em
    T\kern-.1667em\lower.7ex\hbox{E}\kern-.125emX}}
\begin{document}

\title{Hierarchical User Clustering for mmWave-NOMA Systems\\
}

\author{
\IEEEauthorblockN{ Dileepa Marasinghe, Nalin Jayaweera, Nandana Rajatheva and Matti Latva-aho }
\IEEEauthorblockA{\textit{Centre for Wireless Communications,} \\
\textit{Univeristy of Oulu,}\\
Oulu, Finland \\
E-mail: dileepa.marasinghe@oulu.fi,
nalin.jayaweera@oulu.fi,
nandana.rajatheva@oulu.fi, matti.latva-aho@oulu.fi}

}

\maketitle
\begin{abstract}
Non-orthogonal multiple access (NOMA) and mmWave are two complementary technologies that can support the capacity demand that arises in 5G and beyond networks. The increasing number of users are served simultaneously while providing a solution for the scarcity of the bandwidth. In this paper we present a method for clustering the users in a mmWave-NOMA system with the objective of maximizing the sum-rate. An unsupervised machine learning technique, namely, hierarchical clustering is utilized which does the automatic identification of the optimal number of clusters. The simulations prove that the proposed method can maximize the sum-rate of the system while satisfying the minimum QoS for all users without the need of the number of clusters as a prerequisite when compared to other clustering methods such as k-means clustering.
\end{abstract}


\section{Introduction}
With the increasing number of connected devices demanding high bandwidth in 5G and beyond networks, sub-six GHz range fails to support the requirement. MmWave communication was identified as a potential solution\cite{b1} to overcome this by utilizing the enormous bandwidth available in the higher frequencies such as 28 GHz and 60 GHz. However, the characteristics of the mmWave channel are different from the sub-six GHz range. The propagation in the mmWave range is highly directional, suffers from high path loss and has low penetration. However, these can be overcome by the use of beamforming techniques with multiple antennas.

Also, non-orthogonal multiple access (NOMA) gained wide interest as a method of increasing the number of users that can be served simultaneously, thus increasing the spectral efficiency. It has been proved that NOMA outperforms conventional orthogonal multiple access (OMA) methods. Specifically, power domain NOMA where multiplexing is done in the power domain while sharing the same frequency and time is a promising technique to support more users \cite{b2}. In the power domain NOMA principle, superposition coding (SC) is used at the base station and successive interference cancellation (SIC) is employed in the receivers which allow higher throughput\cite{b3}.

These two complementary technologies lay the foundation to mmWave-NOMA systems in the downlink scenario where the highly directional nature of the mmWave channel allows the users who have a correlated channel to be clustered together. Each cluster is served with a mmWave beam and NOMA is applied inside the clustered beam. Such a system presents the problem of grouping the users optimally such that the users in one cluster have highly correlated channels and less correlation with the other clusters. This will lead to reduced interference which will result in utilizing the resources efficiently while maximizing the throughput. The clustering problem with a higher number of users in the system is combinatorial in nature which motivates the use of machine learning techniques to find an optimal solution\cite{b12}. Therefore, in this paper, we mainly focus on the clustering problem presenting an unsupervised machine learning technique, namely the hierarchical clustering, to cluster the users and decide the optimal steering direction of the mmWave beams to serve the cluster.

 \subsection{Prior work}
The problem of clustering the users in NOMA systems has been studied under different scenarios in literature. A cluster head based algorithm for clustering the users using channel gain differences and correlation in downlink multi-user MIMO-NOMA systems is presented in \cite{b5}. Along with user clustering, the optimal power allocation problem for the clusters to support the NOMA principle arises. A scheme for user grouping exploiting the channel gain differences and power allocation for uplink and downlink is investigated in \cite{b6} by forming a sum-rate maximization problem that showed the superiority of NOMA over OMA in performance by system-level simulations. Instead of multi-user clusters, the problem of forming optimal user pairs has been studied in the literature. A greedy search based user pairing method and an iterative method for power allocation are proposed for NOMA systems in \cite{b7} and \cite{b8} with a channel state sorting pairing algorithm for pairing the existing users, and a user difference selecting access for new incoming users.

When considering the mmWave-NOMA systems, a random beamforming technique that uses the correlation of mmWave users' channels is presented in \cite{b9}. The authors propose a randomly generated beam since all the users' channel state information might not be available to the base station to carry out conventional beamforming. This also reduces the overhead of the system in sending the channel state information (CSI) to the base station thus being suitable to fast time-varying situations. Further, in \cite{b10} random beamforming based sum-rate maximization problem is solved by the use of branch and bound approach. Also matching theory and successive convex approximation techniques have been used to solve the problem of power allocation and user scheduling. 

Machine learning-based methods for user clustering for mmWave NOMA systems are described in \cite{b11} and \cite{b12}. User distribution is modeled as a mixture of Gaussians and the expectation-maximization method is employed to tackle the user clustering problem in \cite{b11}. A k-means clustering-based technique for clustering the users with the normalized channel direction and a closed-form power allocation strategy are extensively described in \cite{b12}. However, the k-means clustering algorithm initially needs the number of clusters to run the algorithm and the resulting clustering depends on the random selection of the initial mean points. The user distribution inside the cell is random which makes it difficult to predict the number of clusters initially. In \cite{b12} the authors suggest the use of the elbow or L method for determining the choice of the number of clusters for k-means clustering. However, the downside using that approach with k-means is the requirement to run the k-means algorithm multiple times, varying the number of clusters. We propose the use of hierarchical clustering based method to overcome the initial requirement of having a known number of clusters and discover the number of clusters when running the algorithm which is more appropriate for a random scenario such as the user distribution in a cell.

\begin{figure}[h]
\centering\includegraphics[width=0.8\columnwidth,keepaspectratio]{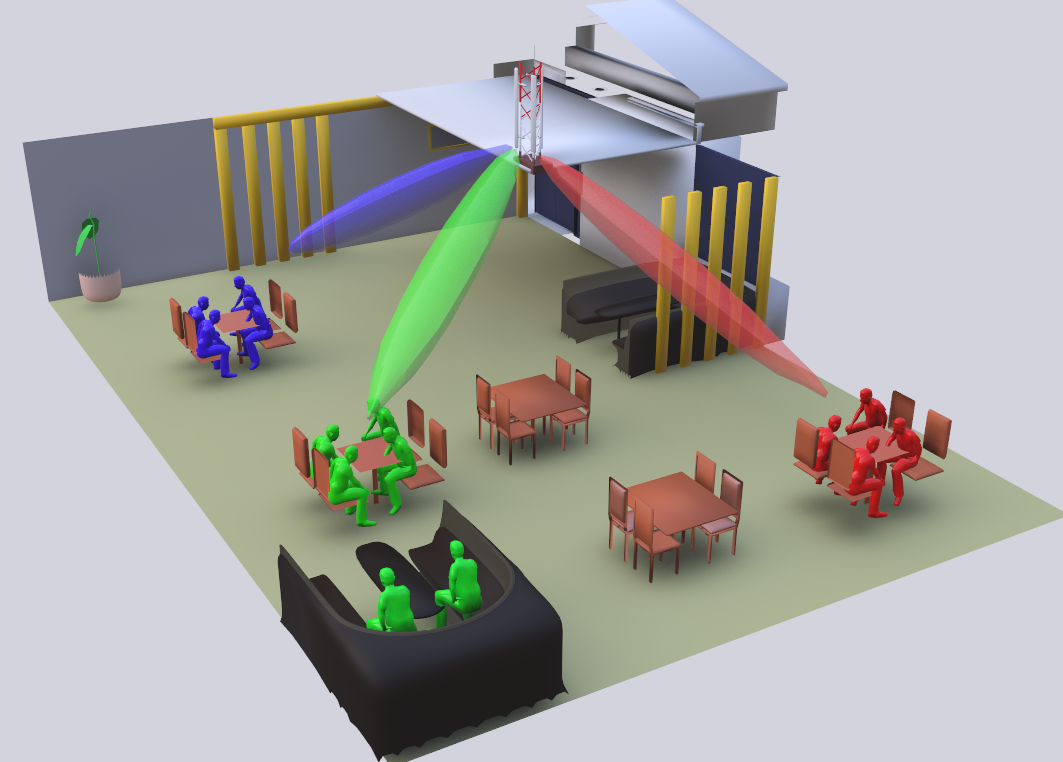}
\caption{An example clustered user distribution scenario in a coffee shop.}
\label{fig1}
\end{figure}

\section{System model}
We consider a single-cell mmWave downlink system that has one base station (BS) equipped with M antennas and N single-antenna users. All the users are served simultaneously and will be clustered into K clusters. Each cluster is served with a single beam while the different clusters are served orthogonally. The users will employ SIC decoding in the clusters following the NOMA principle.

\subsection{User distribution model}

We use a clustered user distribution model as in \cite{b12} which resembles a physically clustered user scenario in an indoor environment such as a low-power small cell in a coffee shop \cite{b13} as in Figure \ref{fig1}. A Poisson cluster process (PCP) is employed to model the user distribution where parent points of the cluster are scattered in a Poisson Point Process (PPP).  The user location $\mathbf{x}_{k,u}$, in the $k^{th}$ cluster relative to the parent point is assumed to be distributed independently and identically (i.i.d) and given by the density $f_{k}(\mathbf{x}_{k,u})$,

\begin{equation*} f_{k}(\mathbf {x}_{k,u}) = \begin{cases} \dfrac {1}{\pi (R_{k})^{2}}, &\mathrm {if}~\|\mathbf {x}_{k,u}\| \leq R_{k},\\ 0, &\mathrm {otherwise}, \end{cases}\tag{1} \label{eq:1}\end{equation*}

where $R_{k}$  is the radius of the uniform disk for cluster \textit{k}. The number of users in a cluster is different from the other clusters which motivates the use of clustering techniques to enhance the sum-rate of the system while ensuring the minimum QoS requirement of all the users.

\subsection{Channel model}

The mmWave channel model in \eqref{eq:2} is used as the channel vector $\mathbf {h}_{u}$ from the base station to a particular user $\mathbf{u}$ which is discussed in \cite{b14} and used in many prior works\cite{b9}\cite{b12}. The channel vector 

\begin{equation*} \mathbf {h}_{u} =\sqrt {M} \frac {\alpha _{u,0}\mathbf {a}(\theta _{u,0})}{ (1+d_{u}^{\eta _{u,LOS}})} + \sum _{l=1}^{L} \sqrt {M} \frac {\alpha _{u,l}\mathbf {a}(\theta _{u,l})}{ (1+d_{u}^{\eta _{u,NLOS}})},\tag{2}\label{eq:2}\end{equation*}

where $L$ is the total number of multipaths, $\mathbf {a}(\theta _{u,l})$ is the steering vector given by
\begin{align*} \mathbf {a}(\theta _{u,l}) = \frac{1}{\sqrt{M}}\begin{bmatrix} 1,e^{-j\pi \theta _{u,l}},\cdots, e^{-j \pi (M-1) \theta _{u,l}} \end{bmatrix}^{T}.\!\!\!\!\! \\ {}\tag{3}\label{eq:3}\end{align*}

Here  $\theta_{u,l} = 2\frac{D}{\lambda}\sin (\phi _{u,l})$ which is the normalized direction where $D \leq \lambda/2$ is the antenna spacing in the BS, $\lambda$ is the wavelength and $\phi _{u,l}\in [0,2\pi]$ is the angle of departure. The complex channel gain of the $l^{th}$ multipath is $\alpha _{u,l} \sim \mathcal {CN}(0,\sigma^2_{u,l})$.  Further, we assume that $\frac{D}{\lambda} =  \frac{1}{2}$ without loss of generality\cite{b9}.

The first term in \eqref{eq:2} corresponds to the line of sight (LOS) path and the summation corresponds to the non-line of sight (NLOS) paths. Moreover $d_{u}$ denotes the distance between the user $\mathbf{u}$ and the BS. The parameters, $\eta _{u,LOS}$ and $\eta _{u,NLOS}$ are the path loss exponents of the  LOS and NLOS paths for user $\mathbf{u}$. The LOS path in a mmWave channel has 20 dB greater gain than the NLOS paths\cite{b1}, which motivates the use of a simplified path model given by the first term in \eqref{eq:2} which considers only the LOS path\cite{b9} as,

\begin{equation*} \mathbf {h}_{u} = \sqrt{M}\frac {\alpha _{u}\mathbf {a}(\theta _{u})}{ (1+d_{u}^{\eta _{u,LOS}})}.\tag{4}\label{eq:4}\end{equation*}
In this work, we assume that the base station has the knowledge of perfect CSI for all users.

\subsection{Signal model}

Let the $p^{th}$ user ($p\in N_k $) in the $k^{th}$ cluster ($k\in K $) be $u_{k,p}$, where $N_k$ is the number of users in the $k^{th}$ cluster. The messages intended  for users $u_{k,p}$, $\forall p\in N_k$ are denoted by $s_{k,p}$ and they are superposed to form the signal  ${X_k}$ in the beam towards the $k^{th}$ cluster by multiplying with the the power splitting factors $\beta_{k,p}$ by ${X_k} = \sum _{p=1}^{N_{k}} \sqrt {\beta _{k,p}} s_{k,p}$ such that $ \sum _{p=1}^{N_k} \beta_{k,p} = 1$.

Let the power allocation to the $k^{th}$ cluster be $p_k$ such that $ \sum _{k=1}^{K} p_k \leq P_T$, where $P_T$ is the total transmission power available in the BS. If the beamforming vector applied in the transmitter is $\mathbf{w_k}$, then the signal vector towards the $k^{th}$ cluster can be expressed as,
$\mathbf{X}_k = \sqrt{p_k}\mathbf{w_k}X_k$. 
Therefore, the received signal to the user $u_{k,p}$ can be expressed as 

\begin{equation*} \mathbf{y}_{k,p} =\mathbf{h}_{k,p}^H\mathbf{X}_k +\mathbf {h}_{k,p}^{H} \sum _{q\neq k} \mathbf {X}_{q} +\mathbf{n}_{k,p}.\tag{5}\label{eq:5}\end{equation*}

The calculation of the optimal power allocation coefficient for $\beta _{k,p}$ is beyond the scope of this work and we use the power allocation method as described in \cite{b12}. In the aforementioned power allocation scheme, the strongest user having the highest channel gain in the cluster is allocated more power while the other users in the cluster are allocated sufficient power for satisfying their minimum QoS constraints.

\section{Agglomerative hierarchical clustering}

Here we introduce the agglomerative hierarchical clustering (AHC) \cite{b16} method applied to cluster the users in the described system. The main advantage of using agglomerative clustering is, specifying the number of clusters is not needed in contrast to a method such as k-means clustering. 
In AHC, a bottom-up approach is employed starting from a set of $U$ vectors and partitioning them into $K$ clusters such that a distortion function is minimized. The required number of clusters $K$ is determined by comparing the merging cost of clusters.
Partitions $R = \{r_1,r_2,\cdots,r_N\}$ define the clustering by assigning the data vectors $u_i$ to cluster $c_k$, which has the same partition $a$, $c_k = \{u_i|r_i = a\}$. Initially, all the data vectors are treated as singleton clusters and clusters are formed by the merging operations $c_k \leftarrow {c_k \cup c_l}$ \cite{b17}.

Since the mmWave-NOMA system considered uses a single beam to communicate to one cluster, users in a cluster should have strong channel correlation\cite{b12}. Cosine similarity between the two channel vectors of two users has been proved to be an efficient metric for determining the similarity between the two users' channels in mmWave systems as follows.

\begin{equation*} \cos (\mathbf {h}_{i},\mathbf {h}_{j}) = \frac {|\mathbf {h}_{i}^{H} \mathbf {h}_{j}|}{\|\mathbf {h}_{i}\| \|\mathbf {h}_{j}\|},\tag{6}\end{equation*}
which can be simplified as,
\begin{align*}
\cos (\mathbf {h}_{i},\mathbf {h}_{j})=&\left\vert \mathbf {a}(\theta _{i})^{H}\mathbf {a}(\theta _{j})\right\vert 
\\=& \frac {1}{M} \left\vert \sum_{l=0}^{M-1}e^{-j\pi l(\theta _{i}-\theta _{j})} \right\vert 
\\=&\frac{1}{M} \left\vert \frac{1-e^{-j\pi M(\theta _{i}-\theta _{j})}}{1-e^{-j\pi (\theta _{i}-\theta _{j})}} \right\vert
\\=&\frac {1}{M}\begin{vmatrix} \frac {\sin \left({\frac {\pi M (\theta _{i}-\theta _{j})}{2} }\right)}{\sin \left({\frac {\pi (\theta _{i}-\theta _{j})}{2} }\right)}\end{vmatrix}  \label{eq:7}\tag{7}\end{align*}

The result in  (\ref{eq:7}) is the absolute of the Dirichlet kernel. When the difference of the normalized directions reaches zero, which means they are in the same normalized direction, the kernel value increases and when the difference increases, it reduces. Therefore, we can use the difference of the normalized directions as a measure of the channel vector similarity. We calculate the centroid $\overline{\theta_k}$ of the $k^{th}$ cluster by averaging the normalized directions of all the users belonging to it as,

\begin{equation*}
    \overline{\theta_k} = \frac{1}{N_k} \sum_{p=1}^{{N_k}} \theta_{k,p}
\end{equation*}

Therefore, the distortion function will be the mean square error given by

\begin{equation*}
    MSE(\overline{\mathbf{\theta}}) = \frac{1}{N} \sum_{k=1}^{{m}}\sum_{p=1}^{{N_k}} {(\theta_{k,p}-\overline{\theta_{k}})^2},
\end{equation*}
where $m$ is the current number of clusters.

Merge cost (MC) of the two clusters $c_k$ and $c_l$ is computed by,

\begin{equation*}
    MC(c_k,c_l) = \frac{N_k N_l}{N_k + N_l} ( \overline{\theta_k} - \overline{\theta_l})^2
\end{equation*}

The clusters to be merged are selected by Ward's method \cite{b18} as,

\begin{equation*}
    k,l = \arg \min_{\underset{i\neq j}{i,j \in [1,m]}} MC(c_i,c_j).
\end{equation*}

\subsection{Choosing the number of clusters for AHC}

\begin{figure}[h]
\includegraphics[width=\columnwidth]{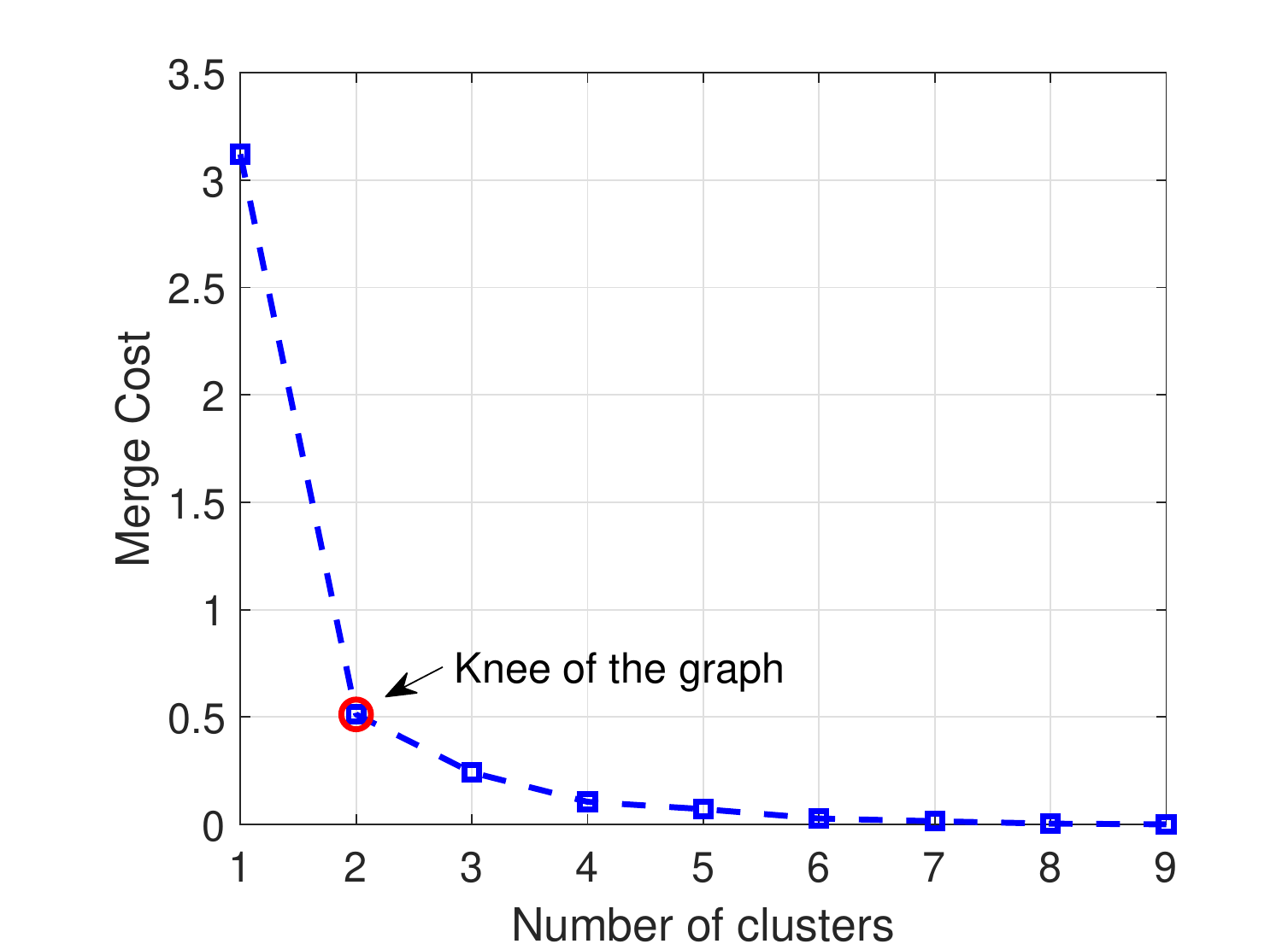}
\caption{An evaluation graph with merging cost as the metric.}
\label{fig2}
\end{figure}

The number of clusters in the AHC algorithm can be decided using the L Method\cite{b19}. The L method identifies the knee of an evaluation graph which has the shape of a letter 'L'. An evaluation graph consists of an evaluation metric plotted against the number of clusters. In this work, the merging cost can be used as the evaluation metric as shown in Figure \ref{fig2}.

This method can be used for other clustering methods such as k-means though the algorithm needs to be re-run completely for different numbers of clusters to obtain an evaluation graph. The advantage of using AHC is, in each step a merging of two clusters is done starting from singleton clusters by considering the merging cost, which can be used as the evaluation metric. When the AHC algorithm is run once, the merging costs are calculated for the whole range of the  number of clusters, ranging from the number of singleton clusters to '1' in the process\cite{b19}.

Consider an evaluation graph plotted with x-axis being the number of clusters ranging from $1$ to $b-1$ where $b$ is the number of singleton clusters. Let us divide the x-axis at $x=c$ yielding right($R_c$) and left sequences($L_c$) such that $L_c = 1\dots c$ and $L_c = c+1\dots b-1$. Now iterate through $c = 2\dots b-3$ with fitting two lines for the coordinates corresponding to $L_c$ and $R_c$ and calculating the root mean square error (RMSE) for the two fits. If the resulting RMSE of the two fits for $L_c$ and $R_c$ are $RMSE(L_c)$ and $RMSE(R_c)$, then the total RMSE can be calculated by,

\begin{equation*}
    RMSE(c) = \frac{c}{b-1} RMSE(L_c) + \frac{b-c}{b-1} RMSE(R_c).
\end{equation*}
The knee of the evaluation graph can be obtained by,

\begin{equation*}
    \hat{c} = \arg \min_{c} RMSE(c). 
\end{equation*}
Then the optimal number of clusters, $K= \hat{c}$.

\section{ Simulations}

The outlined method was evaluated using MATLAB simulations with parameters given in Table \ref{table1}.

\begin{table}[h]

\caption{Simulation parameters}
\begin{tabular}{|l|l|}
\hline
Carrier Frequency                                 & 28 GHz                      \\ \hline
Bandwidth                                         & 2 GHz                       \\ \hline
$\eta_{u,LOS}$ & 2                               \\ \hline
Noise floor                                       & 10 dB                       \\ \hline
Noise power                                       & -174 + 10 log10 (W) +Nf dBm \\ \hline
Boundary radius for parent points                 & 5m                          \\ \hline
Boundary radius for clusters ($R_k$)                 & 1m                          \\ \hline
Number of antennas (M)                                     & 2, 4, 8                          \\ \hline
Minimum QoS constraint for all users              & 0.02                        \\ \hline
\end{tabular}
\label{table1}
\end{table}

\begin{figure}[h]
\includegraphics[width=\columnwidth,keepaspectratio]{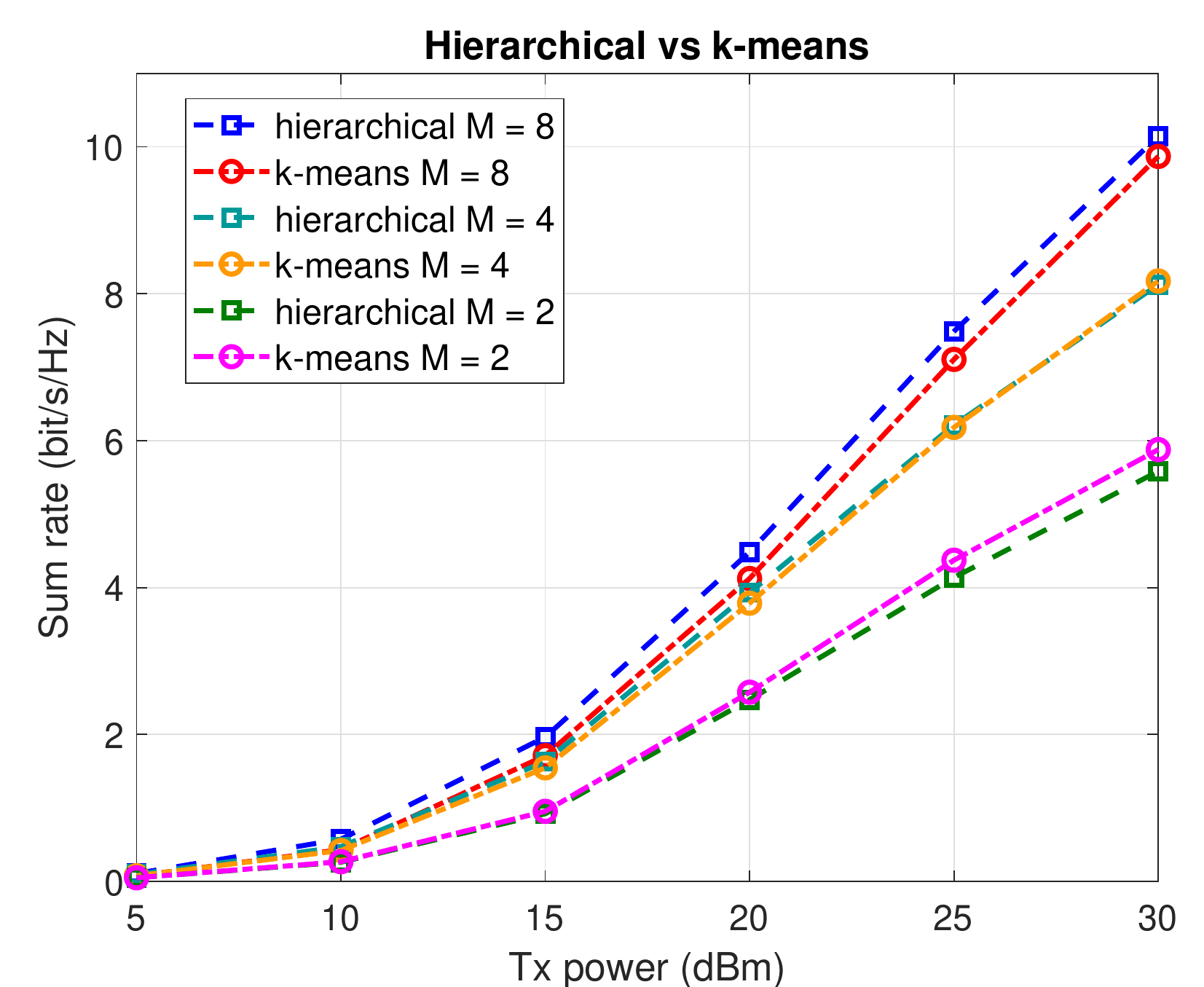}
\caption{Sum rate vs transmit power for $U = 10$.}
\label{fig3}
\end{figure}

A comparison of our method with the k-means method\cite{b12} is carried out with $K=2$ since in \cite{b12}, it is stated that two clusters with k-means results in a higher sum-rate. Our method chooses the optimal number of clusters using the L method described in section III. Figure \ref{fig3} shows the sum-rate of the system plotted against transmit power.

Figure \ref{fig3} shows that when the number of antennas (M) increases the sum-rate increases. The reason for this is when the number of antennas is higher, the beam becomes narrower, which results in a more focused beam towards the cluster center and the interference to other clusters is less. Further, when $M = 8$, our method results in higher sum-rate than the k-means algorithm. When $M = 4$ both methods give similar sum-rate, however, when $M = 2$ k-means performs better. For a less number of antennas, the beam is wider. Since k-means algorithm is run with $K = 2$, when the beam is wider it covers a large area thus performing better compared to our method because our algorithm will decide the optimal number of clusters which can be more than 2.

\begin{figure}[h]
\includegraphics[width=0.9 \columnwidth]{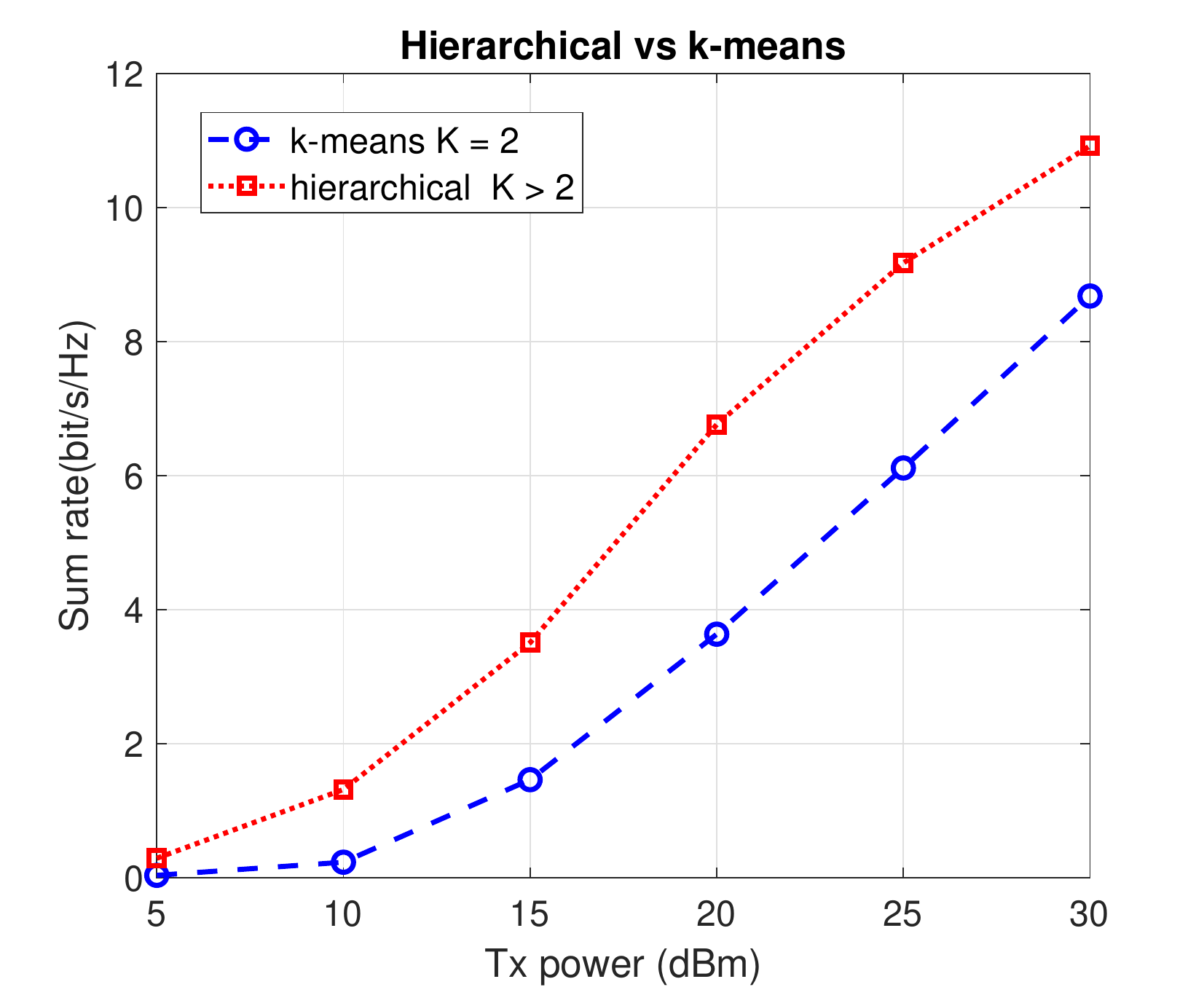}
\caption{AHC resulting in more than 2 clusters with $U = 10$ and $M = 8$.}
\label{fig4}
\end{figure}

\begin{figure}[H]
\includegraphics[width=0.9\columnwidth]{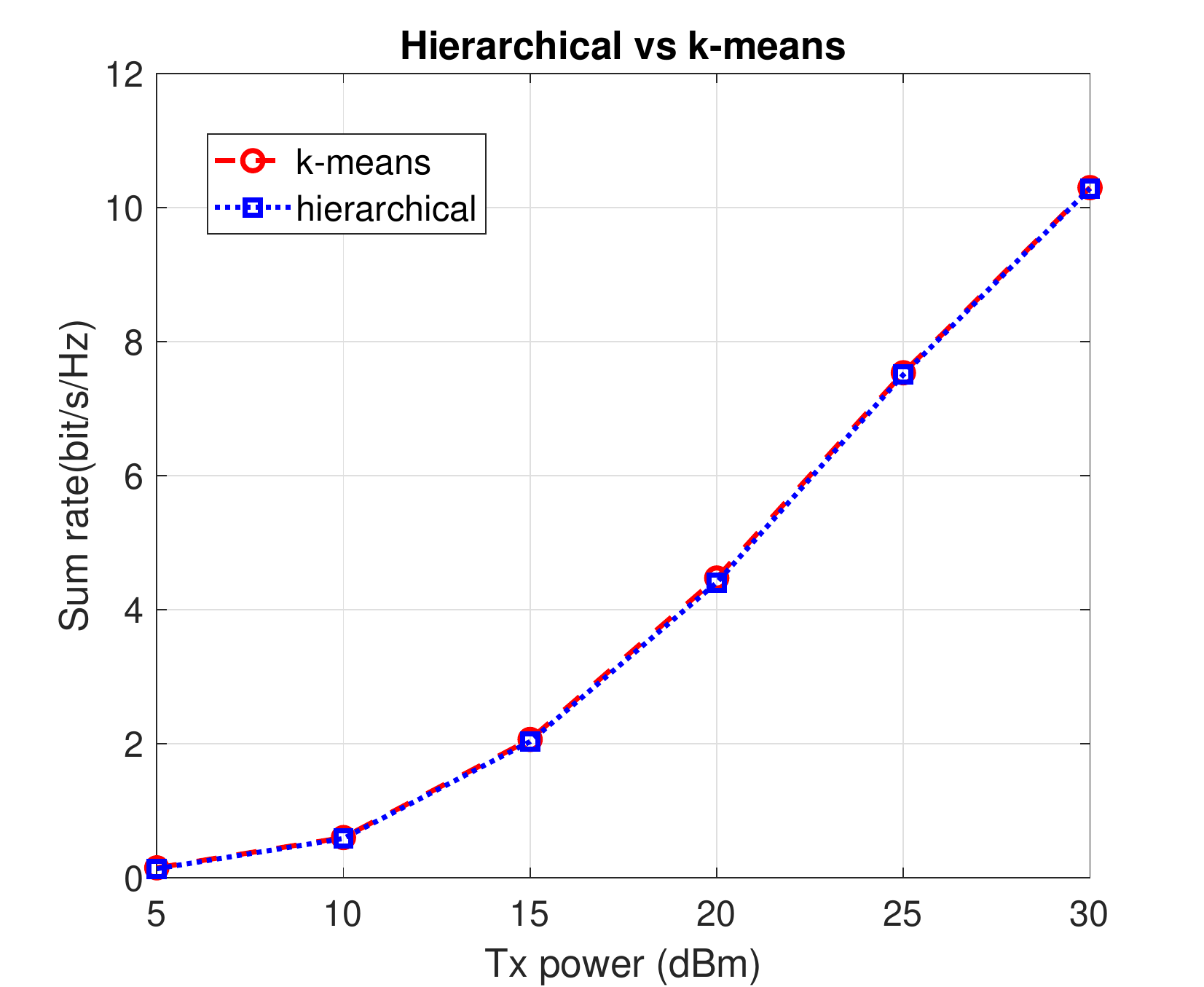}
\caption{K taken from AHC for k-means with $U = 10$ and $M = 8$.}
\label{fig5}
\end{figure}

Figures \ref{fig4} and \ref{fig5} emphasize the importance of choosing an optimal K value. Figure \ref{fig4} is obtained for instances where the AHC clustering results in more than 2 clusters while k-means is set with $K = 2$. Obviously the user distribution consists of more than 2 clusters thus our method results in higher sum-rate.  

Figure \ref{fig5} is obtained by running AHC for a user distribution and identifying the optimal number of clusters first and setting the obtained $K$ value for k-means algorithm. Clearly, both algorithms have similar performance but k-means needs the number of clusters initially. This depicts choosing an optimal K value is necessary. Using our method choosing the number of clusters can be done by running the clustering algorithm once while using k-means requires the clustering to be run multiple times for each number of clusters for the whole user distribution which shows the advantage in using our method.

\section{Conclusion}

In this paper, we discuss the application of agglomerative hierarchical clustering method for mmWave-NOMA systems with the automatic discovery of the optimal number of clusters. The system level simulations prove that the proposed method can maximize the sum-rate of the system while maintaining the minimum QoS for all the users, without the need for the number of clusters as an initial parameter when compared to other clustering methods. Future work can be carried out in considering a multi-cell environment and also for users with mobility.

\end{document}